\documentclass[epj,a4paper]{svjour}
\usepackage{graphicx}
\usepackage{amsfonts}
\usepackage{amsmath}
\usepackage{hyperref}
\usepackage{float}

\begin{document}

\title{Non-extensive statistics, relativistic kinetic theory and fluid
dynamics}
\author{T.\ S.\ Bir\'o\inst{1} \and E.\ Moln\'ar\inst{1,2,3}}
\institute{MTA Wigner Research Centre for Physics, H-1525 Budapest, P.O.Box 49, Hungary 
\and
MTA-DE Particle Physics Research Group, H-4010 Debrecen, P.O.Box 105, 
Hungary 
\and
Frankfurt Institute for Advanced Studies, Ruth-Moufang-Str.\ 1, D-60438
Frankfurt am Main, Germany}

\abstract{
Experimental particle spectra can be successfully described by power-law tailed
energy distributions characteristic to canonical equilibrium distributions
associated to R\'enyi's or Tsallis' entropy formula - over a wide range of energies,
colliding system sizes, and produced hadron sorts. 
In order to derive its evolution one needs a corresponding dynamical description of the system 
which results in such final state observables.
The equations of relativistic fluid dynamics are obtained from a non-extensive Boltzmann
equation consistent with Tsallis' non-extensive $q$-entropy formula. 
The transport coefficients like shear viscosity, bulk viscosity, 
and heat conductivity are evaluate based on a linearized collision integral.
} 
\PACS{ {24.10.Nz}{Hydrodynamic models} \and {05.70.-a} {Entropy in thermodynamics} 
\and{05.20.Dd} {Kinetic theory in statistical mechanics} \and {47.75.+f}{Relativistic fluid dynamics} }

\maketitle

\section{Introduction}

Experimental particle spectra at the Relativistic Heavy-Ion Collider (RHIC)
and at the Large Hadron Collider (LHC) can be successfully described by
power-law tailed energy distributions characteristic to canonical
equilibrium distributions associated to R\'enyi's or Tsallis' entropy formula
- over a wide range of energies, colliding system sizes and produced hadron
sorts \cite%
{EPJA40_Tsallis,EPJA40_Beck,EPJA40_Kaniadakis,EPJA40_Kodama,EPJA40_Wilk,EPJA40_Alberico,EPJA40_Biro,Biro:2008km,Biro:2008er,Urmossy:2011xk,Biro:SQM2011,Urmossy:2012ud,Tang:2008ud,Shao:2009mu,Sikler:HCBM}%
. Such a generalized statistical model can be solicited on the valence quark
level, since several experimental findings on the azimuthal flow component 
$v_2$ \cite{Molnar:2003ff,Fries:2003kq,Zschocke:2011gs} as well as on the
non-extensivity measure $(q-1)$ show quark number scaling 
\cite{Biro:2008er,Urmossy:2011xk,Biro:SQM2011,Urmossy:2012ud}. 
Therefore it is of primary importance to establish a theoretical tool for 
the description of the relativistic flow of such a quark matter.

The non-extensivity effects in general can be viewed as an effective measure
of finite available phase-space \cite{Tsallis_book,Biro_book}, either due to spatial
finiteness of the reaction zone or due to long range entanglement of quark
and gluon strings in the hadron formation process, representing a leading
order deviation between microcanonical and canonical approximations in
traditional terms \cite{Biro_book,Biro:2008vz}. Taken a hadronizing quark
matter with effective, massive and entangled quarks seriously, the $(q-1)$
measure for mesons should be about half, for baryons and antibaryons about
one third of the quark value \cite{Biro:2008er}. It means that if the
power-law tailed observed $p_T$-distribution of hadrons was preformed
practically in the quark phase, then the non-extensivity effects should have
been even more pronounced on the quark than on the hadron level. Therefore
it is unavoidable to test those models, which are based on or at least are
consistent with non-extensive thermodynamics, deriving their equilibrium
state based on the Tsallis' or R\'enyi entropy.

An important step towards this goal is the development of a relativistic
fluid dynamics, which is fully consistent with non-extensivity. Although
initial steps were taken by considering the so-called relativistic perfect 
$q$-hydrodynamics \cite{Osada:2008sw,Osada:2008hs,Osada:2008cn},
the relativistic equations of motion and transport coefficients for a finite 
$q$ have only recently been derived \cite{Biro:2011bq}.

In this paper we present the first full derivation of a consequent fluid
dynamics from a non-extensive relativistic Boltzmann equation (NEBE). This
equation retains the form but generalizes the well known Boltzmann transport
equation, with a single parameter denoted by $q$. We derive the relativistic
fluid-dynamical equations of motion from this non-extensive relativistic
Boltzmann transport equation. Applying well known traditional methods, we
calculate the transport coefficients like heat-conductivity, shear and bulk
viscosity. We also analyse the question which entropy and conserved
Noether four-currents have to be used in a proper, consistent non-extensive
approach. 
Finally we also discuss Grad's method of moments to derive relativistic
causal fluid dynamics in this framework.

\section{Non-extensive statistics, kinetic and fluid dynamical equations}

A non-extensive $q$-generalization of the Boltzmann-Gibbs (BG) entropy was
proposed by C. Tsallis \cite{Tsallis_book,Tsallis} based on the following
formula, 
\begin{equation}
S_{q}\equiv -k_{B}\sum_{i=1}^{W}\left( p_{i}\right) ^{q}\ln _{q}\left(
p_{i}\right) =\frac{k_{B}}{1-q}\left( \sum_{i=1}^{W}\left( p_{i}\right)
^{q}-1\right) ,  \label{Tsallis_entropy}
\end{equation}%
where $k_{B}$ is the Boltzmann constant, $p_{i}$ is the probability of the $%
i $th from a total of $W$ possible microstates, such that $p_{i}\in \left[
0,1\right] $ with the normalization condition $\sum_{i=1}^{W}p_{i}=1$. Here
the $q$-parameter is a real number $q\in \left[ 0,2\right] $, and $\ln _{q}$
is defined as the $q$-logarithmic function, 
\begin{equation}
\ln _{q}\left( x\right) =\frac{x^{1-q}-1}{1-q},  \label{ln_q}
\end{equation}%
for any $x>0$. The inverse of the $q$-logarithmic function is the $q$%
-exponential, 
\begin{equation}
\exp _{q}\left( x\right) \equiv \mathrm{e}_{q}\left( x\right) =\left[
1+\left( 1-q\right) x\right] ^{1/\left( 1-q\right) },  \label{exp_q}
\end{equation}%
for any $x>1/\left( q-1\right) $. These functions return the well known
natural logarithm and exponential functions for $q=1$, that is $\ln
_{q=1}\left( x\right) =\ln \left( x\right) $ and $\exp _{q=1}\left( x\right)
=\exp \left( x\right) $.

For later reference we also define the following dual functions 
\cite{Abe_H_theorem}, 
\begin{eqnarray}
\ln _{q^{\ast }}\left( x\right) &=&\frac{x^{q-1}-1}{q-1}, \\
\mathrm{e}_{q^{\ast }}\left( x\right) &=&\left[ 1+\left( q-1\right) x\right]
^{1/\left( q-1\right) },
\end{eqnarray}%
and 
\begin{eqnarray}
\mathrm{Ln}_{q}\left( x\right) &=&q\left( \frac{x^{q-1}-1}{q-1}\right) , \\
\mathrm{E}_{q}\left( x\right) &=&\left[ 1+\left( \frac{q-1}{q}\right) x%
\right] ^{1/\left( q-1\right) }.
\end{eqnarray}%
One can show using Eqs. (\ref{ln_q},\ref{exp_q}) with $q^{\ast }=2-q$ that $%
\ln _{q^{\ast }}\left( x\right) =\ln _{2-q}\left( x\right) $, $\mathrm{e}%
_{q^{\ast }}\left( x\right) =\mathrm{e}_{2-q}\left( x\right)$. Similarly, $%
\mathrm{E}_{q}\left( x\right) =\mathrm{e}_{q^{\ast }}\left( x/q\right) $, $%
\mathrm{e}_{q}\left( x\right) =\mathrm{E}_{q^{\ast }}\left( q^{\ast
}x\right) $ and $\mathrm{Ln}_{q}\left( x\right) \equiv -q\ln _{q}\left(
1/x\right) =q\ln _{q^{\ast }}\left( x\right) $. These relations are called
dualities and using these functions we can re-write the $q$-generalized 
entropy in two further equivalent forms: 
\begin{equation}
S_{q}\equiv -k_{B}\sum_{i=1}^{W}p_{i}\ln _{q^{\ast }}\left( p_{i}\right) =-%
\frac{k_{B}}{q}\sum_{i=1}^{W}p_{i}\mathrm{Ln}_{q}\left( p_{i}\right) .
\end{equation}%
We note that the R\'enyi formula, $S_{R}=-\frac{k_{B}}{q-1}\ln
\sum_{i=1}^{W}p_{i}^{q}$ is preferred in order to simply fulfill the
additivity property inherent in the zeroth law of thermodynamics \cite%
{BiroVanPRE2011}. 
Since the R\'enyi entropy is proportional to the logarithm of a phase-space 
integral, its local four-current density cannot be
uniquely defined. Consequently, for deriving fluid dynamics and keeping the
linear integration-operator structure of the Boltzmann equation, the Tsallis formula is the one
which provides us a straightforward derivation.

Let us introduce the single particle phase-space distribution function $%
f_{k}=f\left( t,\mathbf{x},k^{0},\mathbf{k}\right)$. 
The distribution function is normalized to the number of particles
in the system, hence in an invariant phase-space volume element 
$\left( \mathbf{x,x} + d\mathbf{x,k,k} + d\mathbf{k}\right) $ 
at time $t$, the number of particles is given by, 
\begin{equation}
d\mathcal{N}\left( t\right) =\ f_{k}d^{3}xd^{3}k,  \label{Nr_particles_1}
\end{equation}
where $x=|\mathbf{x}|$ and $k=|\mathbf{k}|$. From
now on we work with natural units and set $k_{B}=\hbar =c=1$.

The $q$-generalized entropy from Eq. (\ref{Tsallis_entropy}) can also be
expressed with the help of the single particle distribution function.  This leads to
the following relativistic $q$-generalized entropy four-current, 
\begin{equation}
S^{\mu }=-\int dK\ k^{\mu }\left[ \left( f_{k}\right) ^{q}\ln _{q}\left(
f_{k}\right) -f_{k}\right] .  \label{kinetic_S_mu_1}
\end{equation}%
Here $dK = gd^{3}\mathbf{k/}\left( (2\pi )^{3}k^{0}\right)$ is the invariant
momentum-space volume with $g$ denoting the number of internal degrees of
freedom, such as the spin degeneracy. The timelike component of the
relativistic entropy four-current $S^{0}=-\int d^{3}\mathbf{k}\left( g%
\mathbf{/}(2\pi )^{3}\right) \left[ \left( f_{k}\right) ^{q}\ln _{q}\left(
f_{k}\right) -f_{k}\right] $ retains the familiar Tsallis entropy, while the
remaining components define a three vector.

It is imperative to point out that in kinetic theory there are different
definitions of the $q$-entropy four-current in use. These definitions are
not equivalent with each other, and also lead to different thermodynamical
relations as will be discussed later in this paper. 
For example, Lima and Silva et al. \cite{Lima_2001,Lima_2005} applied $S_{S}^{\mu }=-\int \frac{%
d^{3}\mathbf{k}}{k^{0}}\ k^{\mu }f_{k}^{q}\ln _{q}f_{k}$, while Lavagno \cite%
{Lavagno_2002,Lavagno_2009} used $S_{L}^{\mu }=-\int \frac{d^{3}\mathbf{k}}{%
k^{0}}\ k^{\mu }f_{k}^{q}\left[ \ln _{q}f_{k}-1\right] $. 
These alternative
definitions did not explicitly denote the finite size of the elementary
phase-space cell volume $(2 \pi \hbar)^3$. The corrections for quantum
statistics can be added, both for fermions ($z=1$) and for bosons ($z=-1$).
A correct $q$-generalized quantum statistical entropy was proposed by
Cleymans and Worku recently \cite%
{Cleymans:2011ij,Cleymans:2011in,Cleymans:2012ya}, 
\begin{equation}
S_{Q}^{\mu} \! = \!- \! \int \! dK k^{\mu }\left[ f_{k}^{q}\ln _{q}f_{k}+%
\frac{1}{z} \left(1-zf_{k}\right)^{q}\ln _{q}\left( 1-zf_{k}\right) \right] .
\end{equation}%
\linebreak The $z=0$ limit corresponds to Eq. (\ref{kinetic_S_mu_1}) and was
used in this form by Osada and Wilk \cite%
{Osada:2008sw,Osada:2008hs,Osada:2008cn}.

The second law of thermodynamics demands positive entropy production, 
\begin{equation}
\partial _{\mu }S^{\mu }=-\int dK\ln _{q}\left( f_{k}\right) \ \left[ k^{\mu
}\partial _{\mu }\left( f_{k}\right) ^{q}\right] \geq 0,
\label{kinetic_entropy_production_1}
\end{equation}%
This entropy production formula suggests the following kinetic equation,
first introduced by Lavagno \cite{Lavagno_2002}, from now on referred to
as NEBE, 
\begin{equation}
k^{\mu }\partial _{\mu }\tilde{f}_{k}=C\left[ f\right] ,  \label{NEBE_1}
\end{equation}%
where $C\left[ f\right] $ is the collision integral and we use the
notation, 
\begin{equation}
\tilde{f}_{k}=\left( f_{k}\right) ^{q}\ .  \label{f_tilde}
\end{equation}%
This NEBE is postulated with the purpose to study classical statistical
effects due to correlations possibly induced by finite system size, compared
to the characteristic interaction range.

We also stress here that this distribution determines a different number of
particles from Eq. (\ref{Nr_particles_1}), 
\begin{equation}
d\mathcal{\tilde{N}}\left( t\right) = \tilde{f}_{k}d^{3}xd^{3}k\ .
\end{equation}%
Of course this should not mean that the total number of particles in a
system would turn out to be different after integration over the whole
phase-space.

The collision integral, $C\left[ f\right] $, specifies the change in $\tilde{%
f}_{k}$ due to binary collisions among particles with initial incoming
momenta of $k^{\mu }$, $k^{\prime \mu }$ and outgoing final momenta $p^{\mu
} $, $p^{\prime \mu }$. It is constructed with the help of the invariant
transition rate $W_{kk\prime \rightarrow pp\prime }$. To fulfill the 
detailed balance property, the transition rate has to be symmetric
with respect to the sequence of final states $W_{kk\prime \rightarrow
pp\prime }=W_{kk\prime \rightarrow p\prime p} $, as well as symmetric for
time reversed processes $W_{kk\prime \rightarrow pp\prime }=W_{pp\prime
\rightarrow kk\prime }$. The collision integral is formally written as, 
\begin{align}
C\left[ f\right] & =\frac{1}{2}\int dK^{\prime }dPdP^{\prime }W_{kk\prime
\rightarrow pp\prime }  \notag \\
& \times \left( H_{q}\left[ f_{p},f_{p^{\prime }}\right] -H_{q}\left[
f_{k},f_{k^{\prime }}\right] \right) ,  \label{C_f_1}
\end{align}%
where the $q$-generalized version of the assumption of molecular chaos, the
so-called $q$-generalized Stosszahlansatz, is given as in 
\cite{Lavagno_2002,Lavagno_2009} 
\begin{equation}
H_{q}\left[ f_{k},f_{k^{\prime }}\right] =\exp _{q}\left[ \ln _{q}\left(
f_{k}\right) +\ln _{q}\left( f_{k^{\prime }}\right) \right] .  \label{H_q_1}
\end{equation}%
This formula is a postulate for the correlation between two
particles with different momenta before and after the collision. For $q=1$
it returns the familiar collision integral where the number of binary
collisions around space-time coordinates $x^{\mu }$ is proportional to $%
H_{1}\left[ f_{k},f_{k^{\prime }}\right] = f_{k}f_{k^{\prime}}$.

Clearly also other postulates can be made, which all satisfy an H-theorem 
\cite{EPJA40_Biro}. The $H_{q}\left[ f_{k},f_{k^{\prime }}\right]$ in the above formula, in general can be
replaced by any $k k'$-symmetric functional. For example, using 
\begin{align}
\mathcal{H}_{q}\left[ f_{k},f_{k^{\prime }}\right] & =\exp _{q}\left[ \ln
_{q}\left( f_{k}\right) +\ln _{q}\left( f_{k^{\prime }}\right) \right.  
\notag \\
& \left. +\left( 1-q\right) \ln _{q}\left( f_{k}\right) \ln _{q}\left(
f_{k^{\prime }}\right) \right] ,  \label{H_q_factroizable}
\end{align}%
turns out to be the original Boltzmann type of factorization ansatz
neglecting correlations, $\mathcal{H}_{q}\left[ f_{k},f_{k^{\prime }}\right]
=f_{k}f_{k^{\prime }}$  even for $q\neq 1$. This however does not lead automatically to the
classical kinetic theory, just then, if the $k^{\mu }$ four-momenta are
treated additively in the collisions. In this paper we assume that this is
the case.

Note that in the literature not only for the entropy but also for the NEBE
collision kernel, slightly different versions of the kinetic equation and
correlation function exist, involving the $\mathrm{Ln}_{q}$ \cite%
{Abe_H_theorem} or $\ln _{q^{\ast }}$ \cite{Lima_2001,Lima_2005} functions.
We return to discuss these later in the Appendix \ref{Appendix_A}.

The entropy production vanishes in local $q$-equilibrium that is, $\partial
_{\mu }S^{\mu }\equiv -\int dK\ln _{q}\left( f_{0k}\right) C\left[ f_{0}%
\right] =0$. Therefore the distribution function which satisfies $\partial
_{\mu }S^{\mu }\left[ f_{0k}\right] =0$ locally, is the $q$-equilibrium
distribution. This follows assuming that $\ln _{q}\left( f_{0k}\right)
=\alpha _{0} - \beta _{0}^{\mu }k_{\mu }$ hence, 
\begin{equation}
f_{0k}=\exp _{q}\left( \alpha _{0}-\beta _{0}k^{\mu }u_{\mu }\right) .
\label{f_equilibrium_1}
\end{equation}%
Here $\alpha _{0}$ and $\beta_0^{\mu} = \beta _{0}u^{\mu }$ are collision
invariants with $u^{\mu }$ being a four-vector normalized to one, $u^{\mu
}u_{\mu }=1$. These quantities are identified with the inverse temperature $%
\beta _{0}=1/T$, the chemical potential over temperature, $\alpha _{0}=\mu
/T $, and the fluid dynamical flow of matter $u^{\mu }$. The formula in Eq. (%
\ref{f_equilibrium_1}) reduces to the well known J\"{u}ttner \cite%
{Juttner,Juttner_dist_2010} distribution or relativistic Maxwell-Boltzmann
distribution for $q=1$, that is $f_{Jk}=\exp \left( \alpha _{0}-\beta
_{0}k^{\mu }u_{\mu }\right) $.

We introduce the $q$-modified particle four-current and symmetric
energy-momentum tensor, 
\begin{eqnarray}
N^{\mu } &=&\int dK\ k^{\mu }\tilde{f}_{k},  \label{kinetic_N_mu} \\
T^{\mu \nu } &=&\int dK\ k^{\mu }k^{\nu }\tilde{f}_{k}.
\label{kinetic_T_mu_nu}
\end{eqnarray}%
Now, using Eq. (\ref{NEBE_1}) the four-divergences of the above quantities lead
to, 
\begin{eqnarray}
\partial _{\mu }N^{\mu } &\equiv &\int dK\ k^{\mu }\partial _{\mu }\tilde{f}%
_{k}=\int dKC\left[ f\right] ,  \label{kinetic_N_mu_cons} \\
\partial _{\mu }T^{\mu \nu } &\equiv &\int dK\ k^{\nu }k^{\mu }\partial
_{\mu }\tilde{f}_{k}=\int dKk^{\nu }C\left[ f\right] .
\label{kinetic_T_mu_nu_cons}
\end{eqnarray}%
Specifying the so-called collision or summation invariants, $\psi
=\alpha - \beta _{\mu }k^{\mu }$, the conservation of particle four-current
and energy-momentum tensor are implied due to the conservation of particle
number and energy-momentum in individual collisions. These follow from $\int
dK\alpha C\left[ f\right] =0$ and $\int dK\beta _{\mu }k^{\mu }C\left[ f%
\right] =0$ for any solution of the Boltzmann equation, 
\begin{eqnarray}
\partial _{\mu }N^{\mu } &=&0,  \label{N_mu_cons} \\
\partial _{\mu }T^{\mu \nu } &=&0.  \label{T_mu_nu_cons}
\end{eqnarray}%
Having all conserved quantities at hand it is straightforward to prove that
the fundamental relation of equilibrium thermodynamics is given as, 
\begin{equation}
S^{\mu }=-\alpha _{0}N^{\mu }+\beta _{0}T^{\mu \nu }u_{\nu }+\beta
_{0}p_{0}u^{\mu },
\end{equation}%
where $p_{0}$ is the equilibrium pressure. Projecting the above equation
with the local flow velocity, we get the familiar form of the
thermodynamical relation, 
\begin{equation}
s_{0}=-\alpha _{0}n_{0}+\beta _{0}\left( e_{0}+p_{0}\right) .
\end{equation}%
Here $n_{0}=\mathcal{J}_{q\left( 1,0\right) }$ is the particle density, $%
e_{0}=\mathcal{J}_{q\left( 2,0\right) }$ is the energy density and $p_{0}=-%
\mathcal{J}_{q\left( 2,1\right) }$ is related to the particle density as $%
\left. \frac{\partial p_{0}}{\partial \alpha _{0}}\right\vert _{\beta _{0}}=%
\frac{n_{0}}{\beta _{0}}$. We introduced the following $q$-generalized
thermodynamic integrals \cite{Biro:2011bq}: 
\begin{align}
\mathcal{I}_{q\left( i,j\right) }& =\frac{1}{\left( 2j+1\right) !!}\int
dK\left( E_{k}\right) ^{i-2j}\left( \Delta ^{\mu \nu }k_{\mu }k_{\nu
}\right) ^{j}f_{0k},  \label{I_i_j} \\
\mathcal{J}_{q\left( i,j\right) }& =\frac{1}{\left( 2j+1\right) !!}\int
dK\left( E_{k}\right) ^{i-2j}\left( \Delta ^{\mu \nu }k_{\mu }k_{\nu
}\right) ^{j}\left( f_{0k}\right) ^{q},  \label{J_i_j} \\
\mathcal{K}_{q\left( i,j\right) }& =\frac{q}{\left( 2j+1\right) !!}\int
dK\left( E_{k}\right) ^{i-2j}\left( \Delta ^{\mu \nu }k_{\mu }k_{\nu
}\right) ^{j}\left( f_{0k}\right) ^{2q-1},  \label{K_i_j}
\end{align}%
where $i,j\geq 0$ are natural numbers and the double factorial is $\left(
2j+1\right) !!=\left( 2j+1\right) !/\left( j!2^{j}\right) $. Furthermore, we
introduced $\Delta ^{\mu \nu }\equiv g^{\mu \nu }-u^{\mu }u^{\nu }$, to
project arbitrary four-vectors into other four-vectors orthogonal to $u^{\mu
}$. The particle four-momentum is decomposed as $k^{\mu }=E_{k}u^{\mu
}+k^{\left\langle \mu \right\rangle }$, where $E_{k}=k^{\mu }u_{\mu }$ is
the Local Rest Frame (LRF) energy of the particle and $k^{\left\langle \mu
\right\rangle }=\Delta ^{\mu \nu }k_{\nu }$ the LRF momentum. The LRF is
defined by $u_{LRF}^{\mu }=\left( 1,0,0,0\right) $.

It is important to realize that the previously introduced thermodynamical
integrals are consistent with basic relations of classical thermodynamics.
At fixed temperature $\left( f_{0k}\right) ^{q}=\left( \partial
f_{0k}/\partial \alpha _{0}\right) |_{\beta _{0}}$ as well as $q\left(
f_{0k}\right) ^{2q-1}=\left( \partial \left( f_{0k}\right) ^{q}/\partial
\alpha _{0}\right) |_{\beta _{0}} $, so one obtains 
\begin{equation}
\mathcal{J}_{q\left( i,j\right) }=\left( \frac{\partial \mathcal{I}_{q\left(
i,j\right) }}{\partial \alpha _{0}}\right) _{\beta _{0}},\ \mathcal{K}%
_{q\left( i,j\right) }=\left( \frac{\partial \mathcal{J}_{q\left( i,j\right)
}}{\partial \alpha _{0}}\right) _{\beta _{0}}.
\end{equation}%
One shows by partial integration that the following recursive relations hold
in equilibrium: 
\begin{eqnarray}
\mathcal{J}_{q\left( i,j\right) } &=&-\frac{1}{\beta _{0}}\mathcal{I}%
_{q\left( i-1,j-1\right) }+\frac{i-2j}{\beta _{0}}\mathcal{I}_{q\left(
i-1,j\right) },  \label{Jq_nk_rec} \\
\mathcal{K}_{q\left( i,j\right) } &=&-\frac{1}{\beta _{0}}\mathcal{J}%
_{q\left( i-1,j-1\right) }+\frac{i-2j}{\beta _{0}}\mathcal{J}_{q\left(
i-1,j\right) }.  \label{Kq_nk_rec}
\end{eqnarray}%
Furthermore, 
\begin{eqnarray}
\dot{\mathcal{I}}_{q(i,j)} &=&\mathcal{J}_{q\left( i,j\right) }\dot{\alpha}%
_{0}-\mathcal{J}_{q\left( i+1,j\right) }\dot{\beta}_{0}, \\
\dot{\mathcal{J}}_{q(i,j)} &=&\mathcal{K}_{q\left( i,j\right) }\dot{\alpha}%
_{0}-\mathcal{K}_{q\left( i+1,j\right) }\dot{\beta}_{0}.
\end{eqnarray}%
These equations provide the $q$-generalized Gibbs-Duhem relations. They not
only reduce to the standard BG thermodynamics for $q=1$ but at the same time
obey relations consistent with standard thermodynamics \cite%
{Biro:2011bq,Cleymans:2012ya}. Similar recursion relations have been
obtained in framework of higher order generalized
thermodynamics \cite%
{Israel:1979wp,Muronga:2006zx,deGroot_book,Cercignani_book}.

\section{Out of q-equilibrium}

So far we have obtained a local $q$-equilibrium distribution, but yet we are
interested in a more general solution of the NEBE. Assume that 
\begin{equation}
f_{k}=f_{0k}+\delta f_{k},  \label{f_k_non_equilibrium}
\end{equation}%
where $\delta f_{k}$ denotes the deviation from local $q$-equilibrium. It is
customary to surmise $\delta f_{k}\ll f_{0k}$. One should not mix this with
an approximation $(q-1)\ll 1$, as pointed out in 
Refs. \cite{Osada:2008sw,Osada:2008hs,Osada:2008cn}.
This distinction becomes physically important when considering the entropy 
production rate. 
Namely, the small $(q-1)$ approximation can spoil the positivity of the exact
expressions in its leading order. This problem is based on the approximate
behaviour of the basic deformed exponential: While 
\begin{equation}
e_{q}(x)=\exp \left\{ \frac{1}{1-q}\ln \left[ 1+(1-q)x\right] \right\} \geq
0,  \label{DEF_EXP_EXACT}
\end{equation}%
for any real argument, its small $(q-1)$ approximation, 
\begin{equation}
e_{q}(x)=e^{x}\cdot \left( 1-\left( 1-q\right) \frac{x^{2}}{2}+\mathcal{O}%
\left[ \left( q-1\right) ^{2}\right] \right) ,  \label{DEF_EXP_APPROX}
\end{equation}%
will have a negative sign for $|x| > \sqrt{\frac{2}{1-q}}$.
Furthermore, even when $q>1$ would be postulated, in some formulas 
$q^{\ast }=2-q <1$ occurs - as it will be apparent later.
Therefore
such approximations loose the positivity property and should not be used in
describing dissipative phenomena.

The formula (\ref{f_k_non_equilibrium}) can be approximated by a series
expansion around local $q$-equilibrium. As previously discussed, in local $q$%
-equilibrium we may define the collision invariant $\psi _{0k}\equiv
\ln _{q}\left( f_{0k}\right) =\alpha _{0}-\beta _{0}k^{\mu }u_{\mu }$, hence 
\begin{eqnarray}
f_{k}\left( \psi \right) &\equiv &f_{k}\left( \psi _{0k}\right) +\frac{%
\partial f_{k}\left( \psi _{0k}\right) }{\partial \psi _{0k}}\phi _{k}+%
\mathcal{O}\left[ \phi _{k}^{2}\right]  \notag \\
&=&f_{0k}+\left( f_{0k}\right) ^{q}\phi _{k}+\mathcal{O}\left[ \phi _{k}^{2}%
\right] ,
\end{eqnarray}%
where $\phi _{k}=\left( \psi -\psi _{0k}\right) $ denotes the difference
between the collision invariants. Therefore comparing the above result with
Eq. (\ref{f_k_non_equilibrium}) we have to apply 
\begin{equation}
\delta f_{k}=\tilde{f}_{0k}\phi _{k}.  \label{delta_f_k}
\end{equation}%
Because the primary physical quantities are associated with $\left(
f_{0k}\right) ^{q}$ henceforth $\delta \tilde{f}_{k}$ has to be calculated
too. To do this, we expand $\tilde{f}_{k}=\tilde{f}_{0k}+\delta \tilde{f}%
_{k} $ around $\tilde{f}_{0k}$, 
\begin{eqnarray}
\tilde{f}_{k} &\equiv &\left( f_{0k}+\tilde{f}_{0k}\phi _{k}+\mathcal{O}%
\left[ \phi _{k}^{2}\right] \right) ^{q}  \notag \\
&=&\tilde{f}_{0k}+q\left( f_{0k}\right) ^{2q-1}\phi _{k}+\mathcal{O}\left[
\phi _{k}^{2}\right] ,
\end{eqnarray}%
and so we obtain, 
\begin{equation}
\delta \tilde{f}_{k}=q\left( f_{0k}\right) ^{2q-1}\phi _{k}.
\label{delta_f_k_tilde}
\end{equation}

Let us recall the $q$-generalized Stosszahlansatz (\ref{H_q_1}). To further
simplify our discussion we expand it in series, 
\begin{align}
H_{q}\left[ f_{k},f_{k^{\prime }}\right] & =H_{q}\left[ f_{0k},f_{0k^{\prime
}}\right] +\frac{\partial H_{q}\left[ f_{0k},f_{0k^{\prime }}\right] }{%
\partial f_{0k}}\delta f_{k}  \notag \\
& +\frac{\partial H_{q}\left[ f_{0k},f_{0k^{\prime }}\right] }{\partial
f_{0k^{\prime }}}\delta f_{k^{\prime }}+\mathcal{O}\left[ \delta f^{2}\right]
,
\end{align}%
where 
\begin{eqnarray}
\frac{\partial H_{q}\left[ f_{0k},f_{0k^{\prime }}\right] }{\partial f_{0k}}
&=&\left( H_{q}\left[ f_{0k},f_{0k^{\prime }}\right] \right) ^{q}\left(
f_{0k}\right) ^{-q},\  \\
\frac{\partial H_{q}\left[ f_{0k},f_{0k^{\prime }}\right] }{\partial
f_{0k^{\prime }}} &=&\left( H_{q}\left[ f_{0k},f_{0k^{\prime }}\right]
\right) ^{q}\left( f_{0k^{\prime }}\right) ^{-q}.
\end{eqnarray}%
Now, making use of Eq. (\ref{delta_f_k}) we arrive at, 
\begin{equation}
H_{q}\left[ f_{k},f_{k^{\prime }}\right] =H_{q}\left[ f_{0k},f_{0k^{\prime }}%
\right] +\left( H_{q}\left[ f_{0k},f_{0k^{\prime }}\right] \right)
^{q}\left( \phi _{k}+\phi _{k^{\prime }}\right) .
\end{equation}%
In this way the collision integral from Eq. (\ref{C_f_1}) is approximated up
to first order in deviations from $q$-equilibrium as,%
\begin{equation}
C\left[ f\right] =C\left[ f_{0}\right] +C\left[ \delta f\right] .
\end{equation}%
Here $C\left[ f_{0}\right] \propto H_{q}\left[ f_{0p},f_{0p^{\prime }}\right]
-H_{q}\left[ f_{0k},f_{0k^{\prime }}\right] $ vanishes by definition, $H_{q}%
\left[ f_{0p},f_{0p^{\prime }}\right] =H_{q}\left[ f_{0k},f_{0k^{\prime }}%
\right] $. Then $C\left[ f\right] =C\left[ \delta f\right] $ is given by the
integral, 
\begin{align}
C\left[ \delta f\right] & =\frac{1}{2}\int dK^{\prime }dPdP^{\prime
}W_{kk\prime \rightarrow pp\prime }\left( H_{q}\left[ f_{0k},f_{0k^{\prime }}%
\right] \right) ^{q}  \notag \\
& \times \left( \phi _{p}+\phi _{p^{\prime }}-\phi _{k}-\phi _{k^{\prime
}}\right) ,  \label{C_delta_f}
\end{align}%
and so the NEBE (\ref{NEBE_1}) simplifies to, $k^{\mu }\partial _{\mu }%
\tilde{f}_{k}=C\left[ \delta f\right] $.

Using the out-of $q$-equilibrium distribution function, the particle
four-current and the energy-momentum tensor from Eqs. (\ref{kinetic_N_mu},%
\ref{kinetic_T_mu_nu}) take the general form, 
\begin{align}
N^{\mu }& =nu^{\mu }+V^{\mu },  \label{N_mu_q_decomp} \\
T^{\mu \nu }& =eu^{\mu }u^{\nu }-p\Delta ^{\mu \nu }+u^{\mu }W^{\nu }+u^{\nu
}W^{\mu }+\pi ^{\mu \nu }.  \label{T_mu_nu_q_decomp}
\end{align}
We can uniquely identify the fundamental fluid dynamical quantities like the
particle density, energy density and isotropic pressure by following the
matching procedure, 
\begin{eqnarray}
n &\equiv &u_{\mu }N^{\mu }=\int dKE_{k}\tilde{f}_{k}, \\
e &\equiv &u_{\mu }u_{\nu }T^{\mu \nu }=\int dKE_{k}^{2}\tilde{f}_{k}, \\
p &\equiv &-\frac{1}{3}\Delta _{\mu \nu }T^{\mu \nu }=-\frac{1}{3}\int
dK\left( \Delta ^{\alpha \beta }k_{\alpha }k_{\beta }\right) \tilde{f}_{k}.
\end{eqnarray}%
Similarly the particle diffusion and energy-momentum four-currents as well as the
stress tensor are defined as, 
\begin{align}
V^{\mu }& \equiv \Delta _{\alpha }^{\mu }N^{\alpha }=\int dKk^{\left\langle
\mu \right\rangle }\delta \tilde{f}_{k},  \label{def_charge_diffusion} \\
W^{\mu }& \equiv \Delta _{\alpha }^{\mu }u_{\beta }T^{\alpha \beta }=\int
dKE_{k}k^{\left\langle \mu \right\rangle }\delta \tilde{f}_{k},
\label{def_momentum_diffusion} \\
\pi ^{\mu \nu }& \equiv T^{\left\langle \mu \nu \right\rangle }=\int
dKk^{\left\langle \mu \right. }k^{\left. \nu \right\rangle }\delta \tilde{f}%
_{k},  \label{def_stress_tensor}
\end{align}%
where\ $\delta \tilde{f}_{k}=$ $\tilde{f}_{k}-\tilde{f}_{0k}$ and we used
the following constraint $\int dK\left( E_{k}\right) ^{l}\left( \Delta ^{\mu
\nu }k_{\mu }k_{\nu }\right) ^{m}k^{\left\langle \mu _{1}\right. }\ldots
k^{\left. \mu _{n}\right\rangle }\tilde{f}_{0k}=0$ for any $l,n,m$ natural
numbers. Here $T^{\left\langle \mu \nu \right\rangle }=\Delta ^{\mu \nu
\alpha \beta }T_{\alpha \beta }$, with $\Delta ^{\mu \nu \alpha \beta }=%
\frac{1}{2}\left( \Delta ^{\mu \alpha }\Delta ^{\beta \nu }+\Delta ^{\nu
\alpha }\Delta ^{\beta \mu }\right) -\frac{1}{3}\Delta ^{\mu \nu }\Delta
^{\alpha \beta }$ projector, selecting out the traceless, symmetric and orthogonal to 
$u_{\mu}$ part of $T^{\mu \nu }$.

The particle four-current and the energy-momentum tensor calculated from
Eqs. (\ref{kinetic_N_mu},\ref{kinetic_T_mu_nu}) by substituting the local $q$%
-equilibrium distribution function from Eq. (\ref{f_equilibrium_1}), leads
to the so-called \textit{perfect} $q$-fluid decomposition, 
\begin{eqnarray}
N_{0}^{\mu } &=&n_{0}u^{\mu }, \label{perfect_N_mu} \\
T_{0}^{\mu \nu } &=&e_{0}u^{\mu }u^{\nu }-p_{0}\Delta ^{\mu \nu }, \label{perfect_T_mu_nu}
\end{eqnarray}%
where $n_{0}\equiv u_{\mu }N_{0}^{\mu }=\int dKE_{k}\tilde{f}_{0k}$ is the
particle density, $e_{0}\equiv u_{\mu }u_{\nu }T_{0}^{\mu \nu }=\int
dKE_{k}^{2}\tilde{f}_{0k}$ is the energy density and $p_{0}\equiv -\frac{1}{3%
}\Delta _{\mu \nu }T_{0}^{\mu \nu }=-\frac{1}{3}\int dK\left( \Delta
^{\alpha \beta }k_{\alpha }k_{\beta }\right) \tilde{f}_{0k}$ is the pressure
in $q$-equilibrium. The particle density and energy density are usually
assumed to be unchanged from their equilibrium values $n=n_{0}$ and 
$e=e_{0}$, also cited as the Landau matching conditions, 
\begin{eqnarray}
\delta n &\equiv &\int dKE_{k}\delta \tilde{f}_{k}=0,  \label{delta_n} \\
\delta e &\equiv &\int dKE_{k}^{2}\delta \tilde{f}_{k}=0.  \label{delta_e}
\end{eqnarray}%
The isotropic pressure $p$ separates into two parts: $p=p_{0}+\Pi $ where
the bulk viscous pressure becomes, 
\begin{equation}
\Pi =-\frac{1}{3}\int dK\left( \Delta ^{\alpha \beta }k_{\alpha }k_{\beta
}\right) \delta \tilde{f}_{k}.  \label{def_bulk_pressure}
\end{equation}%
The conservation equations for $N_{0}^{\mu }$ and $T_{0}^{\mu \nu }$ turn to
the Euler equations for a perfect $q$-fluid, while Eqs. (\ref{N_mu_cons},\ref%
{T_mu_nu_cons}) together with Eqs. (\ref{N_mu_q_decomp}, \ref%
{T_mu_nu_q_decomp}) constitute the equations of dissipative $q$-fluid
dynamics.

Although we motivated the $q$-fluid dynamical equations form the NEBE, these
constituting equations can also be $\mathit{postulated}$ as the $q$%
-generalized version of the conservation equations based on the 
non-extensive $q$-entropy four-current and the corresponding laws of
thermodynamics. The equations of perfect $q$-fluid dynamics are closed by an
Equation of State (EoS). Without a first order phase transition it implies
a vanishing local entropy production in $q$-equilibrium, $\partial
_{\mu }S_{0}^{\mu }=0$. However, the equations of $q$-fluid dynamics
presented so far in this paper are not closed because the out-of $q$%
-equilibrium $\tilde{f}_{k}$ is unknown. Therefore in this case one needs
additional relations beyond the EoS for completing the system of equations.

\section{The Navier-Stokes equations and transport coefficients}

Taking advantage of the previously derived approximations we reduce the
problem to finding $\delta f_{k}=\tilde{f}_{0k}\phi _{k}$ near local $q$%
-equilibrium. There are several well known approximations to obtain the
deviations from local equilibrium as well as to calculate the transport
coefficients. Here we follow the traditional approach of Chapman and Enskog 
\cite{deGroot_book,Cercignani_book,Chapman_book} and apply this method for the NEBE. 
To this end we note that this method is iterative, and here we take only the first step, 
which results in the well known Navier-Stokes equations of fluid dynamics.

Assuming that in the vicinity of local $q$-equilibrium $\delta \tilde{f}%
_{k}\ll $ $\tilde{f}_{0k}$, the non-equilibrium contributions from the
streaming term are also assumed to vanish $\ k^{\mu }\partial _{\mu }\delta 
\tilde{f}_{k}=0$. Therefore $k^{\mu }\partial _{\mu }\tilde{f}_{k}\simeq $ $%
k^{\mu }\partial _{\mu }\tilde{f}_{0k}$, i.e., 
\begin{align}
k^{\mu }\partial _{\mu }\tilde{f}_{0k}
& =qf_{0k}^{2q-1} \! \left[ -\beta _{0}\frac{\theta }{3}\left( k^{\alpha
}k^{\beta }\Delta _{\alpha \beta }\right) +E_{k}\dot{\alpha}_{0}-E_{k}^{2}%
\dot{\beta}_{0}\right.  \notag \\
& \left. +\left( 1-E_{k}h_{0}^{-1}\right) k^{\left\langle \mu \right\rangle
}\nabla _{\mu }\alpha _{0}-\beta _{0}k^{\left\langle \mu \right. }k^{\left.
\nu \right\rangle }\sigma _{\mu \nu }\right] . \label{CE_lhs}
\end{align}%
Here, $\theta =\nabla _{\mu }u^{\mu }$ is the expansion scalar, $\sigma
^{\mu \nu }=\nabla ^{\left\langle \mu \right. }u^{\left. \nu \right\rangle }$
is the shear-stress tensor and $\nabla _{\mu }\alpha _{0}$ is the gradient
of the chemical potential over temperature. The proper time derivatives $%
\dot{\alpha}_{0}$ and $\dot{\beta}_{0}$ can be expressed as 
\begin{eqnarray}
\dot{\alpha}_{0} &=&n_{0}\mathcal{D}_{q\left( 2,0\right) }^{-1}\left( h_{0}%
\mathcal{K}_{q\left( 2,0\right) }-\mathcal{K}_{q\left( 3,0\right) }\right)
\theta ,  \label{alpha_dot} \\
\dot{\beta}_{0} &=&n_{0}\mathcal{D}_{q\left( 2,0\right) }^{-1}\left( h_{0}%
\mathcal{K}_{q\left( 1,0\right) }-\mathcal{K}_{q\left( 2,0\right) }\right)
\theta ,  \label{beta_dot}
\end{eqnarray}%
where $h_{0}\equiv \left( e_{0}+p_{0}\right) /n_{0}=\mathcal{K}_{q\left(
3,1\right) }/\mathcal{K}_{q\left( 2,1\right) }$ is the enthalpy per particle
and $\mathcal{D}_{q\left( i,j\right) }=\mathcal{K}_{q\left( i-1,j\right) }%
\mathcal{K}_{q\left( i+1,j\right) }-\mathcal{K}_{q\left( i,j\right) }^{2}$.

The main equation we need to solve in the first order Chapman-Enskog method
reduces to, 
\begin{equation}
k^{\mu }\partial _{\mu }\tilde{f}_{0k}=C\left[ \delta f\right] ,
\label{CE_NEBE_eq}
\end{equation}%
where the left-hand-side is given by Eq. (\ref{CE_lhs}), while $C\left[
\delta f\right] $ is given by Eq. (\ref{C_delta_f}). Inspecting this
integro-differential equation, we search for its solution using the ansatz 
\cite{Cercignani_book}, 
\begin{align}
\phi _{k}& =\left( \varphi _{0}+\varphi _{1}E_{\mathbf{k}}+\varphi _{2}E_{%
\mathbf{k}}^{2}\right) \theta +\left( \varphi _{3}+\varphi _{4}E_{\mathbf{k}%
}\right) k^{\left\langle \mu \right\rangle }\nabla _{\mu }\alpha _{0}  \notag
\\
& +\varphi _{5}k^{\left\langle \mu \right. }k^{\left. \nu \right\rangle
}\sigma _{\mu \nu },  \label{phi_k}
\end{align}%
where the $\varphi _{i}$'s are yet unknown coefficients. In an earlier work 
\cite{Biro:2011bq} we calculated the transport coefficients based on the
Anderson-Witting model \cite{Anderson_Witting} where $\phi _{k}$ was given
directly by Eq. (\ref{CE_lhs}) using the so-called relaxation time
approximation, $C\left[\delta f\right] =-E_{k}\left( \tilde{f}_{k}-\tilde{f}%
_{0k}\right) /\tau _{C}$, with $\tau _{C} $ being the mean time between
collisions. Similar calculations were also done in the non-relativistic limit 
by Bezerra et al. \cite{Lima_trans_coeffs_2003} using an alternative NEBE 
from Refs. \cite{Lima_2001,Lima_2005}. 

Here we proceed with the calculation of the $\varphi _{i}$ coefficients.
Using Eqs. (\ref{delta_n},\ref{delta_e}) and the definition of the bulk
viscous pressure from Eq. (\ref{def_bulk_pressure}) together with Eq. (\ref%
{delta_f_k_tilde}) we get three coupled equations for the first three
coefficients: 
\begin{align}
\delta n& \equiv q\int dKE_{k}\left( f_{0k}\right) ^{2q-1}\phi _{k}  \notag
\\
& =\left( \varphi _{0}\mathcal{K}_{q\left( 1,0\right) }+\varphi _{1}\mathcal{%
K}_{q\left( 2,0\right) }+\varphi _{2}\mathcal{K}_{q\left( 3,0\right)
}\right) \theta , \\
\delta e& \equiv q\int dKE_{k}^{2}\left( f_{0k}\right) ^{2q-1}\phi _{k} 
\notag \\
& =\left( \varphi _{0}\mathcal{K}_{q\left( 2,0\right) }+\varphi _{1}\mathcal{%
K}_{q\left( 3,0\right) }+\varphi _{2}\mathcal{K}_{q\left( 4,0\right)
}\right) \theta , \\
\Pi & \equiv -\frac{q}{3}\int dK\left( \Delta ^{\alpha \beta }k_{\alpha
}k_{\beta }\right) \left( f_{0k}\right) ^{2q-1}\phi _{k}  \notag \\
& =-\left( \varphi _{0}\mathcal{K}_{q\left( 2,1\right) }+\varphi _{1}%
\mathcal{K}_{q\left( 3,1\right) }+\varphi _{2}\mathcal{K}_{q\left(
4,1\right) }\right) \theta .
\end{align}%
For $\delta n=\delta e=0$ from the matching conditions (\ref{delta_n},\ref%
{delta_e}), one concludes that 
\begin{eqnarray}
\varphi _{0} &=&-\frac{\Pi }{\theta }\frac{\mathcal{D}_{q\left( 3,0\right) }%
}{D_{\Pi }}, \\
\varphi _{1} &=&-\frac{\Pi }{\theta }\frac{\left( \mathcal{K}_{q\left(
2,0\right) }\mathcal{K}_{q\left( 3,0\right) }-\mathcal{K}_{q\left(
1,0\right) }\mathcal{K}_{q\left( 4,0\right) }\right) }{D_{\Pi }}, \\
\varphi _{2} &=&-\frac{\Pi }{\theta }\frac{\mathcal{D}_{q\left( 2,0\right) }%
}{D_{\Pi }},  \label{CE_phi_2}
\end{eqnarray}%
where the denominator is given by 
\begin{align}
D_{\Pi }& =\mathcal{K}_{q\left( 2,1\right) }\mathcal{D}_{q\left( 3,0\right)
}+\mathcal{K}_{q\left( 4,1\right) }\mathcal{D}_{q\left( 2,0\right) }  \notag
\\
& +\mathcal{K}_{q\left( 3,1\right) }\left( \mathcal{K}_{q\left( 2,0\right) }%
\mathcal{K}_{q\left( 3,0\right) }-\mathcal{K}_{q\left( 1,0\right) }\mathcal{K%
}_{q\left( 4,0\right) }\right) .
\end{align}%
It is important to note that the matching conditions are not 
unique and other choices may be made \cite%
{Stewart,Osada:2011gx,Osada:2012yp}. In general one may assume that $\delta
n\neq \delta e\neq 0$, so the coefficients $\varphi _{i}\left( \delta
n,\delta e,\Pi \right) $ are functions of all scalar non-equilibrium
corrections. However, as we shall see it later, the constraints to the NEBE
only supplies $\varphi _{2}$, the other two coefficients are left
undetermined. Therefore, in general we need additional input to specify the
non-equilibrium scalar corrections, and in this paper we work
with the matching conditions from Eqs. (\ref{delta_n},\ref{delta_e}).

Similarly, the definitions of the particle and energy-momentum
diffusion four-currents and the stress tensor lead to,
\begin{eqnarray}
V^{\mu } &\equiv &q\int dKk^{\left\langle \mu \right\rangle }\left(
f_{0k}\right) ^{2q-1}\phi _{k}  \notag \\
&=&\left( \varphi _{3}\mathcal{K}_{q\left( 2,1\right) }+\varphi _{4}\mathcal{%
K}_{q\left( 3,1\right) }\right) \nabla ^{\mu }\alpha _{0},  \label{CE_V_mu}
\\
W^{\mu } &\equiv &q\int dKE_{k}k^{\left\langle \mu \right\rangle }\left(
f_{0k}\right) ^{2q-1}\phi _{k}  \notag \\
&=&\left( \varphi _{3}\mathcal{K}_{q\left( 3,1\right) }+\varphi _{4}\mathcal{%
K}_{q\left( 4,1\right) }\right) \nabla ^{\mu }\alpha _{0},  \label{CE_W_mu}
\\
\pi ^{\mu \nu } &\equiv &q\int dKk^{\left\langle \mu \right. }k^{\left. \nu
\right\rangle }\left( f_{0k}\right) ^{2q-1}\phi _{k}  \notag \\
&=&2\varphi _{5}\mathcal{K}_{q\left( 4,2\right) }\sigma ^{\mu \nu }.
\label{CE_pi_mu_nu}
\end{eqnarray}%
Here we used, $k^{\left\langle \mu \right. }k^{\left. \nu \right\rangle
}=k^{\left\langle \mu \right\rangle }k^{\left\langle \nu \right\rangle }-%
\frac{1}{3}\left( \Delta ^{\alpha \beta }k_{\alpha }k_{\beta }\right) \Delta
^{\mu \nu }$, and the general orthogonality relation, 
\begin{eqnarray}
&&\int dK\text{ }\mathrm{F}(E_{\mathbf{k}})\ k^{\left\langle \mu _{1}\right.
}...k^{\left. \mu _{m}\right\rangle }k_{\left\langle \nu _{1}\right.
}...k_{\left. \nu _{n}\right\rangle }  \notag \\
&=&\frac{m!\,\delta _{mn}}{\left( 2m+1\right) !!}\Delta _{\nu _{1}...\nu
_{m}}^{\mu _{1}...\mu _{m}}\int dK\ \mathrm{F}(E_{\mathbf{k}})\ \left(
\Delta ^{\alpha \beta }k_{\alpha }k_{\beta }\right) ^{m},\ \ \ \ 
\end{eqnarray}%
with $\mathrm{F}(E_{\mathbf{k}})$ being a function of energy \cite%
{deGroot_book}. 

These results for the dissipative quantities present the
relativistic Navier-Stokes relations, usually written in the more familiar
form, 
\begin{eqnarray}
\Pi &=&-\zeta \theta ,  \label{NS_bulk} \\
q^{\mu } &=&-\frac{\kappa }{h_{0}\beta _{0}^{2}}\nabla ^{\mu }\alpha _{0},
\label{NS_heat} \\
\pi ^{\mu \nu } &=&2\eta \sigma ^{\mu \nu }.  \label{NS_shear}
\end{eqnarray}%
The relativistic counterpart of the Stokes result, $\Pi =-\zeta
\theta $, relates the bulk viscous pressure to the expansion rate by
introducing the coefficient of bulk viscosity $\zeta $. The
Fourier-Navier-Stokes law in Eq. (\ref{NS_heat}) relates the heat-flow $%
q^{\mu }=W^{\mu }-h_{0}V^{\mu }$ to the temperature gradients in the system, 
$\nabla ^{\mu }\alpha _{0}=-h_{0}T^{-2}\left( \nabla ^{\mu }T-T\dot{u}^{\mu
}\right) $ with the coefficient of heat- or thermal-conductivity $\kappa $.
Finally the Newton-Navier-Stokes relation between the stress and shear is
given in Eq. (\ref{NS_shear}) with the coefficient of shear viscosity $\eta $.

The $\varphi _{i}$ coefficients are obtained inserting $\phi _{k}$ from Eq. (%
\ref{phi_k}) into the collision integral (\ref{C_delta_f}) and matching the
gradients on both sides. The collision integral leads to, 
\begin{align}
C\left[ \delta f\right]\! & = \! \frac{1}{2} \! \int dK^{\prime
}dPdP^{\prime }W_{kk\prime \rightarrow pp\prime }\left( H_{q}\left[
f_{0k},f_{0k^{\prime }}\right] \right) ^{q}  \notag \\
& \times \! \left[ \varphi _{2}\left( E_{p}^{2}+E_{p^{\prime
}}^{2}-E_{k}^{2}-E_{k^{\prime }}^{2}\right) \theta \right.  \notag \\
& \left. +\varphi _{4} \! \left( E_{p}p^{\left\langle \mu \right\rangle
}+E_{p^{\prime }}p^{\prime \left\langle \mu \right\rangle
}-E_{k}k^{\left\langle \mu \right\rangle }-E_{k^{\prime }}k^{\prime
\left\langle \mu \right\rangle }\right) \nabla _{\mu }\alpha _{0}\right. 
\notag \\
& \left. +\varphi _{5}\! \left( p^{\left\langle \mu \right. }p^{\left. \nu
\right\rangle }+p^{\prime \left\langle \mu \right. }p^{\prime \left. \nu
\right\rangle }-k^{\left\langle \mu \right. }k^{\left. \nu \right\rangle
}-k^{\prime \left\langle \mu \right. }k^{\prime \left. \nu \right\rangle
}\right) \sigma _{\mu \nu }\right] ,  \label{C_delta_f_final}
\end{align}%
where the contributions vanish identically for $\varphi _{0}$, $\varphi _{1}$
and $\varphi _{3}$ due to the conservation of charge, energy and momenta in
binary collisions.

For later purposes we introduce the $X^{\mu \nu \alpha \beta }$ tensor, symmetric upon the
interchange of indexes $\left( \mu ,\nu \right) $ as well as $\left( \alpha ,\beta
\right) $, while traceless for the latter, $X^{\mu \nu \alpha \beta
}g_{\alpha \beta }=0$: 
\begin{align}
X^{\mu \nu \alpha \beta }& =\frac{1}{2}\int dKdK^{\prime }dPdP^{\prime
}W_{kk\prime \rightarrow pp\prime }\left( H_{q}\left[ f_{0k},f_{0k^{\prime }}%
\right] \right) ^{q}  \notag \\
& \times k^{\mu }k^{\nu }\left( p^{\alpha }p^{\beta }+p^{\prime \alpha
}p^{\prime \beta }-k^{\alpha }k^{\beta }-k^{\prime \alpha }k^{\prime \beta
}\right) .  \label{X_tensor}
\end{align}%
Furthermore, one can show that $X^{\mu \nu \alpha \beta }$ is generally
decomposed as,%
\begin{align}
X^{\mu \nu \alpha \beta }& =\left( X_{1}u^{\mu }u^{\nu }+X_{2}\Delta ^{\mu
\nu }\right) \left( u^{\alpha }u^{\beta }-\frac{1}{3}\Delta ^{\alpha \beta
}\right)  \notag \\
& +4X_{3}u^{\left( \mu \right. }\Delta ^{\left. \nu \right) \left( \alpha
\right. }u^{\left. \beta \right) }+X_{4}\Delta ^{\mu \nu \alpha \beta },
\label{X_4rank_tens}
\end{align}%
with the coefficients, 
\begin{align}
X_{1}& \equiv X^{\mu \nu \alpha \beta }u_{\mu }u_{\nu }u_{\alpha }u_{\beta
}=-X^{\mu \nu \alpha \beta }u_{\mu }u_{\nu }\Delta _{\alpha \beta },
\label{X1_tens} \\
X_{2}& \equiv \frac{1}{3}X^{\mu \nu \alpha \beta }\Delta _{\mu \nu
}u_{\alpha }u_{\beta }=-\frac{1}{3}X^{\mu \nu \alpha \beta }\Delta _{\mu \nu
}\Delta _{\alpha \beta },  \label{X2_tens} \\
X_{3}& \equiv \frac{1}{3}X^{\mu \nu \alpha \beta }u_{\left( \mu \right.
}\Delta _{\left. \nu \right) \left( \alpha \right. }u_{\left. \beta \right)
}=\frac{1}{3}X^{\mu \nu \alpha \beta }u_{\mu }\Delta _{\nu \alpha }u_{\beta
},  \label{X3_tens} \\
X_{4}& =\frac{1}{5}X^{\mu \nu \alpha \beta }\Delta _{\mu \nu \alpha \beta }.
\label{X4_tens}
\end{align}

Applying these results and integrating Eq. (\ref{CE_NEBE_eq}) we get three
equations which are proportional to the different Navier-Stokes type
gradients occurring in Eqs. (\ref{NS_bulk}-\ref{NS_shear}). For example,
collecting terms proportional to $\sigma ^{\alpha \beta }$ leads to, 
\begin{eqnarray}
&&-\beta _{0}\int dKk^{\mu }k^{\nu }k_{\left\langle \alpha \right.
}k_{\left. \beta \right\rangle }\left( qf_{0k}^{2q-1}\right)  \notag \\
&=&\frac{\varphi _{5}}{2}\int dKdK^{\prime }dPdP^{\prime }W_{kk\prime
\rightarrow pp\prime }\left( H_{q}\left[ f_{0k},f_{0k^{\prime }}\right]
\right) ^{q}  \notag \\
&&\times k^{\mu }k^{\nu }\left( p_{\left\langle \alpha \right. }p_{\left.
\beta \right\rangle }+p_{\left\langle \alpha \right. }^{\prime }p_{\left.
\beta \right\rangle }^{\prime }-k_{\left\langle \alpha \right. }k_{\left.
\beta \right\rangle }-k_{\left\langle \alpha \right. }^{\prime }k_{\left.
\beta \right\rangle }^{\prime }\right) ,\ \ 
\end{eqnarray}%
which briefly reads as, $-2\beta _{0}\mathcal{K}_{q\left( 4,2\right) }\Delta
^{\mu \nu \alpha \beta }=\varphi _{5}X^{\mu \nu \alpha \beta}$. From this we
easily get that 
\begin{equation}
\varphi _{5}=-2\beta _{0}\frac{\mathcal{K}_{q\left( 4,2\right) }}{X_{4}}.
\end{equation}%
Replacing this result into Eq. (\ref{CE_pi_mu_nu}) and comparing it to Eq. (%
\ref{NS_shear}) leads to the coefficient of shear viscosity, 
\begin{equation}
\eta \equiv \varphi _{5}\mathcal{K}_{q\left( 4,2\right) }=-2\beta _{0}\frac{%
\mathcal{K}_{q\left( 4,2\right) }^{2}}{X_{4}}.
\end{equation}

Similarly, the coefficient of heat-conductivity can be calculated, by
matching the gradient $\nabla ^{\mu }\alpha _{0}\ $on both sides of the
integrated NEBE. This leads to, 
\begin{equation}
\varphi _{4}=-\frac{\mathcal{K}_{q\left( 3,1\right) }-h_{0}^{-1}\mathcal{K}%
_{q\left( 4,1\right) }}{X_{3}}.
\end{equation}%
The $\varphi _{3}$ coefficient can be found only after fixing the LR frame.
Here we have at least two choices. We may choose $V^{\mu }=0$, which
corresponds to Eckart's definition \cite{Eckart:1940te}, and hence the fluid
dynamical flow velocity is fixed to the flow of conserved particles, $u^{\mu
}=N^{\mu }/\sqrt{N^{\mu }N_{\mu }}$. Another choice due to Landau and
Lifshitz \cite{Landau_book} defines the flow as the time-like eigenvector of
the energy-momentum tensor, $u^{\mu }=T^{\mu \nu }u_{\nu }/\sqrt{T^{\mu
\alpha }u_{\alpha }T_{\mu \beta }u^{\beta }}$. This choice is equivalent to $%
W^{\mu }=0$. Choosing the Landau frame, one arrives at 
\begin{equation}
\varphi _{3}=-\varphi _{4}\frac{\mathcal{K}_{q\left( 4,1\right) }}{\mathcal{K%
}_{q\left( 3,1\right) }},
\end{equation}%
and thus the coefficient of heat conductivity becomes 
\begin{equation}
\kappa =\varphi _{4}h_{0}^{2}\beta _{0}^{2}\frac{\mathcal{D}_{q\left(
3,1\right) }}{\mathcal{K}_{q\left( 3,1\right) }}.
\end{equation}

Finally, the coefficient of bulk viscosity can be expressed with the help of
Eqs. (\ref{alpha_dot}, \ref{beta_dot}): 
\begin{align}
\varphi _{2}& =-\frac{\mathcal{D}_{q\left( 2,0\right) }^{-1}}{X_{1}}\left[
\beta _{0}\mathcal{K}_{q\left( 4,1\right) }\mathcal{D}_{q\left( 2,0\right)
}-n_{0}\mathcal{D}_{q\left( 3,0\right) }\right.  \notag \\
& \left. -n_{0}h_{0}\left( \mathcal{K}_{q\left( 2,0\right) }\mathcal{K}%
_{q\left( 3,0\right) }-\mathcal{K}_{q\left( 1,0\right) }\mathcal{K}_{q\left(
4,0\right) }\right) \right] ,
\end{align}%
and hence from Eqs. (\ref{CE_phi_2}) and (\ref{NS_bulk}) we obtain, 
\begin{equation}
\zeta =\varphi _{2}\frac{D_{\Pi }}{\mathcal{D}_{q\left( 2,0\right) }}.
\end{equation}

These results formally resemble their classical counterparts within BG
statistics. An explicit $q$-dependence occurs in the thermodynamic integrals only 
while the coefficients $X_1,X_2,X_3$ and $X_4$ differ in their arguments.

In the followings we attempt to simplify the
collisional integral.
Recall the $H_{q}\left[ f_{k},f_{k^{\prime }}\right] $ functional from Eq. (%
\ref{H_q_1}) and calculate it in equilibrium getting 
\begin{eqnarray}
\left( H_{q}\left[ f_{0k},f_{0k^{\prime }}\right] \right) ^{q} &\equiv
&\left( \exp _{q}\left[ \ln _{q}\left( f_{0k}\right) +\ln _{q}\left(
f_{0k^{\prime }}\right) \right] \right) ^{q}  \notag \\
&=&\left[ 1+\left( 1-q\right) \left( \psi _{0k}+\psi _{0k^{\prime }}\right) %
\right] ^{\frac{q}{1-q}}.
\end{eqnarray}%
Now making use of Eq. (\ref{exp_q}) one concludes that 
\begin{equation}
H_{q}\left[ f_{0k},f_{0k^{\prime }}\right] =\exp _{q}\left( \psi _{0k}+\psi
_{0k^{\prime }}\right) .
\end{equation}%
The above formula for $q=1$ leads to the classical result, $H_{1}\left[
f_{0k},f_{0k^{\prime }}\right] =f_{Jk}f_{Jk^{\prime }}$ with $f_{Jk}\equiv
\exp \left( \psi _{0k}\right) =\exp \left( \alpha _{0}-\beta _{0}k^{\mu
}u_{\mu }\right) $. Substituting this result into Eq. (\ref{X_tensor}) we
arrive at, 
\begin{align}
X^{\mu \nu \alpha \beta }& =\frac{1}{2}\int dKdK^{\prime }dPdP^{\prime
}W_{kk\prime \rightarrow pp\prime }\left( \exp _{q}\left( \psi _{0k}+\psi
_{0k^{\prime }}\right) \right) ^{q}  \notag \\
& \times k^{\mu }k^{\nu }\left( p^{\alpha }p^{\beta }+p^{\prime \alpha
}p^{\prime \beta }-k^{\alpha }k^{\beta }-k^{\prime \alpha }k^{\prime \beta
}\right) .  \label{X_tensor_last}
\end{align}
This integral can be rewritten with the help of a Mandelstam invariant $%
s\equiv \left( k^{\mu }+k^{\prime \mu }\right) ^{2}=\left( p^{\mu
}+p^{\prime \mu }\right) ^{2}$. So the transition rate is given as \cite%
{deGroot_book}, 
\begin{equation}
W_{kk\prime \rightarrow pp\prime }=\left( 2\pi \right) ^{6}s\sigma \left(
s,\theta _{CM}\right) \delta ^{4}\left( k^{\mu }+k^{\prime \mu }-p^{\mu
}-p^{\prime \mu }\right) .  \label{Transition_rate}
\end{equation}%
Here $\sigma \left( s,\theta _{CM}\right) $ is the differential cross
section, $\theta _{CM}$ is the scattering angle in the center of momentum frame, 
while the $\delta $-function represents the energy-momentum
conservation in binary collisions.

Substituting the transition rate into Eq. (\ref{X_tensor_last}), we get the
following expression in the center of mass frame for the $p$-dependent
integral: 
\begin{align}
\mathcal{I}_{p}\left( \sigma _{T},s,m\right) & \equiv \frac{1}{2 }\int 
\frac{d^{3}\mathbf{p}}{\left( p^{0}\right) ^{2}}s\ \sigma \left( s,\theta
_{CM}\right) \delta \left( \sqrt{s}-2p^{0}\right) ,  \notag \\
& =\frac{\sigma _{T}\left( s\right) }{2}\sqrt{s\left( s-4m^{2}\right) }.
\end{align}%
Here we defined the total cross-section $\sigma _{T}\left( s\right) $ as the
integral over the differential cross-section and the solid angle, $d\Omega
=\int_{0}^{2\pi }d\varphi \int_{0}^{\pi }\sin \theta _{CM}d\theta _{CM}$,
hence 
\begin{equation}
\sigma _{T}\left( s\right) =\frac{1}{2}\int_{0}^{\pi }d\Omega \ \sigma
\left( s,\theta _{CM}\right) .  \label{cross_section}
\end{equation}%
Finally the $X^{\mu \nu \alpha \beta }$ tensor can be given as, 
\begin{align}
X^{\mu \nu \alpha \beta }\!& =\!\frac{1}{2}\int dKdK^{\prime }\ \mathcal{I}%
_{p}\left( \sigma _{T},s,m\right) \left( \exp _{q}\left( \psi _{0k}+\psi
_{0k^{\prime }}\right) \right) ^{q}  \notag \\
& \!\times k^{\mu }k^{\nu }\Bigg\{\frac{1}{2}\left[ P^{\alpha }P^{\beta }-\frac{%
\left( s-4m^{2}\right) }{3}\left( g^{\alpha \beta }-\frac{P^{\alpha }P^{\beta }}{s}%
\right) \right]  \notag \\
& \!-\left( k^{\alpha }k^{\beta }+k^{\prime \alpha }k^{\prime \beta }\right) %
\Bigg\},
\end{align}%
where we introduced the total momentum $P^{\mu }\equiv k^{\mu }+k^{\prime \mu
}=p^{\mu }+p^{\prime \mu }$. This integral, contrary to the classical
examples is not easy to handle analytically due to the fact that only $\exp
_{q}\left[ x+y+\left( 1-q\right) xy\right] =\exp _{q}\left( x\right) \exp
_{q}\left( y\right) $, hence the $k$ integrals can not be factorized into
thermodynamical integrals.

As already shown the thermodynamical quantities and transport coefficients change 
with $q$-parameter. The fact that the collision integral does not factorize, 
signals a direct $q$-dependence and the effect of the correlations included in the NEBE.
Therefore it also follows that the original Boltzmann equation ($q=1$) neglects 
any kind of memory effects and two particle correlations, hence the transport 
coefficients calculated from such an equation also lack correlations.

\section{Conclusions and outlook}

In this work we derived the $q$-generalized versions of the classical
Navier-Stokes equations of relativistic dissipative fluid dynamics from a $q$%
-generalized Boltzmann transport equation. These equations were found based
on the Chapman-Enskog method.

We showed that starting from a $q$-generalized transport equation, it is justified to apply 
standard methods to calculate transport coefficients. These
calculations lead to relations for all transport coefficients formally
similar to those one would obtain using the traditional Boltzmann-Gibbs
distributions. The main difference is contained in the recursive rules for
the thermodynamic integrals.

However, unlike in traditional fluid dynamics, the tensorial collision
kernel $X^{\mu \nu \alpha \beta }$ does not factorize into product of
simple thermodynamical integrals. This remarkable property is a consequence
of the $q$-deformed exponential function describing the extended local $q$%
-equilibrium distribution.

We also add that the derivation presented in this work is applicable to slightly different 
kinetic equations as discussed in Appendix \ref{Appendix_A}, 
as well as to the one suggested by 
G.~Kaniadakis \cite{EPJA40_Kaniadakis,Kaniadakis_2001,Kaniadakis_2010_EPL}.
Furthermore, we also show in Appendix \ref{Grad} that one obtains causal 
fluid dynamics from the NEBE by using Grad's method moments.

These methods applied to the $q$-generalized Boltzmann equation extend 
the applicability of dissipative fluid dynamics for $q\neq 1$ by including
long range interactions and correlations, but for example the resulting 
Navier-Stokes equations are still parabolic, hence problems 
related to acausality are not solved by introducing a non-extensivity 
parameter $q$ in this framework. 
Therefore this means that the causality problem is not rooted in the non-extensivity of entropy, but 
the entropy should be extended to include dissipative quantities which is well 
known from irreversible thermodynamic theories \cite{Muller:1999in,IS}.

On the other hand, we remark that Osada and Wilk in Refs. \cite{Osada:2008sw,Osada:2008hs,Osada:2008cn} 
associated the perfect $q$-hydrodynamics 
with the classical Navier-Stokes equations using a so-called 
non-extensive/dissipative correspondence (NexDC).
This approach makes a direct correspondence between the purely $q$-dependent 
conserved quantities, Eqs. (\ref{perfect_N_mu}, \ref{perfect_T_mu_nu}) 
for $q\neq 1$, and the general dissipative structure as given 
in Eqs. (\ref{N_mu_q_decomp}, \ref{T_mu_nu_q_decomp}) for $q=1$.
Therefore the dissipative quantities are identified with differences 
given by perfect $q$-fluid dynamical quantities for $q\neq 1$ 
and classical dissipative fluid dynamics for $q=1$.

This novel method may reveal the correlations induced by long range effects
that in general contribute to dissipation and entropy production. 
However our approach goes beyond by generalizing the fluid dynamical equations 
to out of $q$-equilibrium states, hence in principle one can directly estimate 
the $q$-parameter without conjecturing a NexDC.

\begin{acknowledgement}
The authors acknowledge valuable discussions with J.~Cleymans, G.~S.~Denicol, T.~F\"ul\"op, 
A.~Muronga, H.~Niemi, and P.~V\'an. 
We also thank T.~Osada and to an anonymous referee for constructive comments and criticism.
This work 
was supported by the National Development Agency NF\"U under the
Hungarian National Scientific Fund contract OTKA K68108 and the bilateral 
South-African Hungarian project TeT10-1-2011-0061 (ZA-15/2009). 
E.~M. was supported by the National Development Agency NF\"U contract
OTKA/NF\"U 81655.
\end{acknowledgement}

\appendix

\section{Appendix}
\subsection{Remarks on a different NEBE}
\label{Appendix_A}

Here we briefly discuss different NEBEs advocated by some authors. The
existence of different kinetic equations shows that there is an ambiguity
about the correct equation of motion, although they are consistent with the
same $q$-generalized entropy.

Let us recall the entropy four-current from Eq. (\ref{kinetic_S_mu_1}) and
rewrite it in equivalent forms: 
\begin{eqnarray}
S^{\mu } &\equiv &-\int dK\ k^{\mu }\left[ f_{k}\ln _{q^{\ast }}\left(
f_{k}\right) -f_{k}\right] ,  \notag \\
&=&-\frac{1}{q}\int dK\ k^{\mu }\left[ f_{k}\mathrm{Ln}_{q}\left(
f_{k}\right) -qf_{k}\right] .  \label{kinetic_S_mu_2}
\end{eqnarray}%
The requirement of positive entropy production turns out to be,%
\begin{equation}
\partial _{\mu }S^{\mu }\equiv-q\int dK\ln _{q^{\ast }}\left( f_{k}\right) 
\left[ k^{\mu }\partial _{\mu }f_{k}\right] \geq 0,
\end{equation}%
and 
\begin{equation}
\partial _{\mu }S^{\mu } = -\int dK\mathrm{Ln}_{q}\left( f_{k}\right) \left[
k^{\mu }\partial _{\mu }f_{k}\right] \geq 0.
\label{kinetic_entropy_production_2}
\end{equation}%
The above formulas suggest a different version of the $q$-generalized
Boltzmann equation. This alternative NEBE can be written as, (denoted with
hats to avoid confusion) 
\begin{equation}
k^{\mu }\partial _{\mu }\hat{f}_{k}=\hat{C}\left[ \hat{f}\right] .
\label{NEBE_2}
\end{equation}%
Here the collision integral is 
\begin{align}
\hat{C}\left[ \hat{f}\right] & =\frac{1}{2}\int dK^{\prime }dPdP^{\prime
}W_{kk\prime \rightarrow pp\prime }  \notag \\
& \times \left( \hat{H}_{q}\left[ \hat{f}_{p},\hat{f}_{p^{\prime }}\right] -%
\hat{H}_{q}\left[ \hat{f}_{k},\hat{f}_{k^{\prime }}\right] \right) ,
\label{C_f_2}
\end{align}%
and 
\begin{equation}
\hat{H}_{q}\left[ \hat{f}_{k},\hat{f}_{k^{\prime }}\right] =\exp _{q}\left[ 
\mathrm{Ln}_{q}\left( \hat{f}_{k}\right) +\mathrm{Ln}_{q}\left( \hat{f}%
_{k^{\prime }}\right) \right] .  \label{H_q_2}
\end{equation}%
The outcome of the H-theorem as shown by Abe \cite{Abe_H_theorem}, is
equivalent to $\partial _{\mu }S^{\mu }\equiv -\int dK\mathrm{Ln}_{q}\left(
f_{0k}\right) \hat{C}\left[ f\right] =0$, whence the collision invariant is $%
\hat{\psi}_{0k}=\mathrm{Ln}_{q}\left( \hat{f}_{0k}\right) $, leading to 
\begin{equation}
\hat{f}_{0k}\equiv \mathrm{E}_{q}\left( \alpha _{0}-\beta _{0}k^{\mu }u_{\mu
}\right) =\exp _{q^{\ast }}\left( \frac{\alpha _{0}-\beta _{0}k^{\mu }u_{\mu
}}{q}\right) \,.  \label{f_equilibrium_2}
\end{equation}%
We further note that Lima and Silva et al. \cite{Lima_2001,Lima_2005} used a
slightly different $q$-generalized Stosszahlansatz given as $H_{q}^{\ast }\left[ f_{k}^{\ast
},f_{k^{\prime }}^{\ast }\right] =\exp _{q}\left[ \ln _{q^{\ast }}\left(
f_{k}^{\ast }\right) +\ln _{q^{\ast }}\left( f_{k^{\prime }}^{\ast }\right) %
\right] $. Actually, this can be achieved by the duality transformation, $\ln
_{q^{\ast }}\left( x\right) =\frac{1}{q}\mathrm{Ln}_{q}\left( x\right) $,
and so their equilibrium distribution function becomes,%
\begin{equation}
f_{0k}^{\ast }=\mathrm{e}_{q^{\ast }}\left( \alpha _{0}-\beta _{0}k^{\mu
}u_{\mu }\right) .  \label{f_equilibrium_3}
\end{equation}
It is clear that the different NEBE's lead to different stationary
solutions. However the cited solution of Abe as well as of Lima and Silva et
al., consequently leads to different thermodynamical relations than ours.

All these different $q$-equilibrium distributions can be used to define
different particle four-currents and symmetric energy-momentum tensors. 
For example, denote $N^{\mu }\left( \hat{f}_{0k}\right) =\hat{N}_{0}^{\mu
}$, $T^{\mu \nu }\left( \hat{f}_{0k}\right) =\hat{T}_{0}^{\mu \nu }$ and $%
S^{\mu }\left( \hat{f}_{0k}\right) =\hat{S}_{0}^{\mu }$, whence by using Eq.
(\ref{kinetic_S_mu_2}) together with the corresponding stationary solution (%
\ref{f_equilibrium_2}), we obtain%
\begin{eqnarray}
\hat{S}_{0}^{\mu } \!&\equiv & \! -\frac{\alpha _{0}}{q}\int dK k^{\mu }\hat{f}%
_{0k}+\frac{\beta _{0}}{q}\int dK k^{\mu }E_{k}\hat{f}_{0k}+\int dK k^{\mu }%
\hat{f}_{0k}  \notag \\
\!&=&\!-\frac{\alpha _{0}}{q}\hat{N}_{0}^{\mu }+\frac{\beta _{0}}{q}\hat{T}%
_{0}^{\mu \nu }u_{\nu }+\hat{N}_{0}^{\mu }.
\end{eqnarray}%
Projecting the above equation and introducing $\hat{\alpha}_{0}=\alpha
_{0}/q\ $ and $\hat{\beta}_{0}=\beta _{0}/q$ we get, $\hat{s}_{0}=-\hat{%
\alpha}_{0}\hat{n}_{0}+\hat{\beta}_{0}\hat{e}_{0}+\hat{n}_{0}$.
Correspondingly one has to define, the following thermodynamic integrals,
similar to Eqs. (\ref{I_i_j},\ref{J_i_j},\ref{K_i_j}), 
\begin{align}
\hat{I}_{q\left( i,j\right) }& = \frac{1}{\left( 2j+1\right) !!}\int
dK\left( E_{k}\right) ^{i-2j}\left( \Delta ^{\mu \nu }k_{\mu }k_{\nu
}\right) ^{j}\left( \hat{f}_{0k}\right) ^{q}, \\
\hat{J}_{q\left( i,j\right) }& = \frac{1}{\left( 2j+1\right) !!}\int
dK\left( E_{k}\right) ^{i-2j}\left( \Delta ^{\mu \nu }k_{\mu }k_{\nu
}\right) ^{j}\hat{f}_{0k}, \\
\hat{K}_{q\left( i,j\right) }& = \frac{q^{-1}}{\left( 2j+1\right) !!}%
\int dK\left( E_{k}\right) ^{i-2j}\left( \Delta ^{\mu \nu }k_{\mu }k_{\nu
}\right) ^{j}\left( \hat{f}_{0k}\right) ^{2-q}.
\end{align}%
Now we rewrite the thermodynamic relation as, 
\begin{equation}
\hat{s}_{0}=-\hat{\alpha}_{0}\hat{J}_{q\left( 1,0\right) }+\hat{\beta}_{0}%
\hat{J}_{q\left( 2,0\right) }+\hat{J}_{q\left( 1,0\right) }.
\label{thermo_2}
\end{equation}%
However at fixed temperature $\hat{f}_{0k}=\left( \partial \left( \hat{f}%
_{0k} \right)^{q}/\partial \alpha _{0}\right) |_{\beta _{0}}$ and 
$q^{-1}\left(\hat{f}_{0k} \right)^{2-q}= \left( \partial \hat{f}%
_{0k}/\partial \alpha _{0}\right) |_{\beta _{0}}$ hence%
\begin{equation}
\hat{J}_{q\left( i,j\right) }=\left( \frac{\partial \hat{I}_{q\left(
i,j\right) }}{\partial \alpha _{0}}\right) _{\beta _{0}}, \
\hat{K}_{q\left( i,j\right) }=\left( \frac{\partial \hat{J}_{q\left(
i,j\right) }}{\partial \alpha _{0}}\right) _{\beta _{0}},
\end{equation}%
while 
\begin{equation}
\hat{J}_{q\left( i,j\right) }=-\frac{1}{\beta _{0}}\hat{I}_{q\left(
i-1,j-1\right) }+\frac{i-2j}{\beta _{0}}\hat{I}_{q\left( i-1,j\right) }. 
\end{equation}%
These relations actually lead to an equilibrium pressure given as, $%
\hat{p}_{0}\equiv -\hat{J}_{q\left( 2,1\right) }=\beta _{0}^{-1}\hat{I}%
_{q\left( 1,0\right) }$ and so the last term from Eq. (\ref{thermo_2})
cannot be identified with the pressure times inverse temperature $\hat{\beta_0} \hat{p}%
_{0} \neq \hat{J}_{q\left( 1,0\right) }$. Therefore the equilibrium state
defined by Eq. (\ref{f_equilibrium_2}) and Eq. (\ref%
{f_equilibrium_3}), are not consistent with the ideal gas EOS in its classical form,
while the one utilized in Eq. (\ref{f_equilibrium_1}) is,  
$\beta_0 p_0 = \mathcal{I}_{q \left( 1,0\right)}$. This does
not necessarily mean that the NEBE\ in Eq. (\ref{NEBE_2}) or its stationary solutions
are ruled out, but one has to keep in mind that they lead to weird
thermodynamic relations.

Similarly, as previously presented we can define the deviations from
equilibrium as, 
\begin{equation}
\hat{f}_{k}=\hat{f}_{0k}+\frac{1}{q}\left( \hat{f}_{0k}\right) ^{2-q}\hat{\phi}
_{k}+\mathcal{O}\left[ \hat{\phi} _{k}^{2}\right] ,  \notag
\end{equation}%
where $\hat{\phi}_{k} =\left( \hat{\psi} - \hat{\psi}_{0k}\right) $ 
and so $\delta \hat{f}_{k}=\frac{1}{q}\left( \hat{f}_{0k}\right) ^{2-q}\hat{\phi}
_{k}$. The collision integral also simplifies and formally corresponds to
Eq. (\ref{C_delta_f}), 
\begin{align}
\hat{C}\left[ \delta f\right] & =\frac{1}{2}\int dK^{\prime }dPdP^{\prime
}W_{kk\prime \rightarrow pp\prime }\left( \hat{H}_{q}\left[ \hat{f}_{0k},%
\hat{f}_{0k^{\prime }}\right] \right) ^{q} \notag \\
& \times \left( \hat{\phi} _{p}+ \hat{\phi} _{p^{\prime }}-\hat{\phi} _{k}- \hat{\phi} _{k^{\prime}}\right) .
\end{align}

Therefore independently of the underlying NEBE the transport coefficients
can be calculated by any classical method of choice. 
However as we have shown, differences arise in the thermodynamical 
integrals and the EoS.

\subsection{Grad's method of moments}
\label{Grad}

Here we discuss an alternative method to calculate the transport
coefficients. This method, originally due to Grad \cite{Grad} leads not only
slightly different transport coefficients but also to different equations of motion for the
dissipative quantities. This is due to the different choice of parameters
for the non-equilibrium distribution function and the particularities of the
method, see for example Refs. \cite{Israel:1979wp,deGroot_book,Cercignani_book,Grad,Denicol:2010xn,Betz:2010cx,Denicol:2012cn,Denicol:2012_new}.
for more details.

Here we follow Refs.\cite{Denicol:2012cn,Denicol:2012_new} and introduce the
following irreducible tensor moment for the deviation form local $q$%
-equilibrium, 
\begin{equation}
\tilde{\rho}_{r}^{\left\langle \mu _{1}\ldots \mu _{\ell }\right\rangle
}=\int dKE_{k}^{r}k^{\left\langle \mu _{1}\right. }\ldots k^{\left. \mu
_{\ell }\right\rangle }\delta \tilde{f}_{k},  \label{gen_moment}
\end{equation}%
where the index $\ell $ indicates the rank of the tensor such that $\ell =0$
corresponds to the scalar $\tilde{\rho}_{r}$, with power $r$ of the energy $%
E_{k}$. Similarly one can introduce irreducible moments involving the
solution of the other NEBE from Eq. (\ref{NEBE_2}), i.e., $\hat{\rho}%
_{r}^{\left\langle \mu _{1}\ldots \mu _{\ell }\right\rangle }=\int
dKE_{k}^{r}k^{\left\langle \mu _{1}\right. }\ldots k^{\left. \mu _{\ell
}\right\rangle }\delta \hat{f}_{k}$. The generalized irreducible moments
from Eq. (\ref{gen_moment}) are identified with the dissipative quantities 
via the following relations,
\begin{eqnarray}
\ \delta n &\equiv &\tilde{\rho}_{\left( 1\right) }=0,\ \delta e\equiv 
\tilde{\rho}_{\left( 2\right) }=0,\Pi =-\frac{m^{2}}{3}\tilde{\rho}_{\left(
0\right) },  \notag \\
\ V^{\mu } &=&\tilde{\rho}_{\left( 0\right) }^{\mu },\ W^{\mu }=\tilde{\rho}%
_{\left( 1\right) }^{\mu },\ \pi ^{\mu \nu }=\tilde{\rho}_{\left( 0\right)
}^{\mu \nu }.
\end{eqnarray}

Now, rewriting the NEBE\ (\ref{NEBE_1}) in the following form, 
\begin{align}
\frac{d}{d\tau }\delta \tilde{f}_{k}& =-\frac{d}{d\tau }\tilde{f}%
_{0k}-E_{k}^{-1}k_{\nu }\nabla ^{\nu }\tilde{f}_{0k}  \notag \\
& -E_{k}^{-1}k_{\nu }\nabla ^{\nu }\delta \tilde{f}_{k}+E_{k}^{-1}C\left[
\delta f\right] \;,  \label{linBoltz}
\end{align}%
we obtain exact equations for $\frac{d}{d\tau }\tilde{\rho}_{r}^{\left\langle
\mu _{1}\cdots \mu _{\ell }\right\rangle }$. This method is presented and
analysed in great detail in Refs. \cite{Denicol:2012cn,Denicol:2012_new}.

The main results of Grad's method are the so-called relaxation equations,
which determine the time evolution of $\tilde{\rho}_{r}^{\left\langle \mu _{1}\ldots \mu _{\ell }\right\rangle}$ hence also of $\Pi $, $q^{\mu }$, and $\pi ^{\mu \nu}$. 
The relaxation of the dissipative quantities towards their Navier-Stokes values 
is given with time scales given by the corresponding relaxation
times $\tau _{\Pi }$, $\tau _{q}$ and $\tau _{\pi }$. The calculation of the
collision integral actually involves the very same procedure as presented in
the previous sections, while the relaxation equations are also given in exactly 
the same form as in the classical case. Once again the $q \ne 1$ modifications are
embedded in the thermodynamical and collision integrals.



\end{document}